\documentclass[pre, letterpaper, amsmath, amssymb, preprintnumbers, showpacs, superscriptaddress, twocolumn, floatfix]{revtex4-1}
\usepackage[pdftex]{graphicx}
\usepackage{amsmath,amssymb,amsfonts,amscd,amsfonts,hyperref,txfonts}

\begin{document}
\title{A characterization of quantum chaos by two-point correlation functions}

\preprint{Brown-HET-1786}

\author{Hrant Gharibyan}
\affiliation{Stanford Institute for Theoretical Physics, Stanford University, Stanford, CA 94305, USA}
\author{Masanori Hanada}
\affiliation{Department of Mathematics, University of Surrey, Guildford, Surrey, GU2 7XH, UK}
\affiliation{School of Physics and Astronomy, and STAG Research Centre, University of Southampton, Southampton, SO17 1BJ, UK}
\affiliation{Department of Physics, Brown University, 182 Hope Street, Providence, RI 02912, USA}
\author{Brian Swingle}
\affiliation{Condensed Matter Theory Center, Maryland Center for Fundamental Physics,
Joint Center for
Quantum Information and Computer Science, and Department of Physics,
University of Maryland,
College Park MD 20742, USA}
\author{Masaki Tezuka}
\affiliation{Department of Physics, Kyoto University, Kyoto 606-8502, Japan}

\begin{abstract}
We propose a characterization of quantum many-body chaos: given a collection of simple operators, the set of all possible pair-correlations between these operators can be organized into a matrix with random-matrix-like spectrum. This approach is particularly useful for locally interacting systems, which do not generically show exponential Lyapunov growth of out-of-time-ordered correlators. We demonstrate the validity of this characterization by numerically studying the Sachdev-Ye-Kitaev model and a one-dimensional spin chain with random magnetic field (XXZ model).
\end{abstract}

\maketitle

\section{Introduction}
How do we characterize quantum chaos? Among a wide variety of different approaches (see Ref.~\onlinecite{2016AdPhy..65..239D} for a review), two rather different criteria are currently in wide use. The first one is random-matrix-like universality of the energy spectrum \cite{PhysRevLett.52.1,haake2013quantum}: a given quantum system is chaotic in this sense if the energy spectrum is described by Gaussian Random Matrix Theory, which we simply denote by RMT \cite{wigner1993characteristic,doi:10.1063/1.1703773,mehta2004random}.
The second one is sensitivity to initial conditions: a given quantum system is chaotic in this sense if it exhibits exponential Lyapunov growth of a small perturbation
as probed by an out-of-time-order correlation function (OTOC) \cite{larkin1969quasiclassical,Almheiri:2013hfa}. OTOCs are closely related to Loschmidt echoes which also probe chaos~\cite{2012arXiv1206.6348G}.

There are several unsatisfactory features regarding these criteria. First, it is unclear how the two criteria are related. Second, the connection of the quantum criteria to the characterizations of classical chaos are unclear. One might expect that sensitivity to initial conditions can characterize both classical and quantum chaos, but there is a problem for local quantum systems. In the classical theory, the initial perturbation can be taken arbitrarily small in the mathematical sense, and the exponential growth can continue forever.
On the other hand, in a quantum system the perturbation cannot be arbitrarily small due to the uncertainty principle,
and local quantum systems do not generally show exponential growth except in special limits~\cite{2018PhRvX...8b1014N,2018PhRvX...8b1013V,2018arXiv180200801X,2018arXiv180505376X,Khemani2018vellyap}~\footnote{In the context of holography, the large-$N$ limit of a gauge theory, a vector model, or the SYK model are often considered. In these examples, $N$ corresponds to the number of the internal degrees of freedom. These internal interactions can be regarded as highly non-local in the sense that all of them interact directly with each other. Large $N$ also plays the role of small $\hbar$, giving a kind of semi-classical limit. In these cases, the exponential growth of OTOC is a good indicator of chaos.}. Hence, the characterization based on the early growth of OTOCs does not work for generic local quantum systems.

In a previous paper \cite{Gharibyan:2018fax}, we generalized the above single chaos exponent to define a spectrum of quantum Lyapunov exponents. Based on calculations in the Sachdev-Ye-Kitaev (SYK) model and a spin chain (XXZ) model, we proposed that the Lyapunov exponents so defined exhibit a universal behavior:
the Lyapunov spectrum agrees with RMT when the system is chaotic. This characterization of quantum chaos circumvented the problem of lack of exponential growth in generic local systems, since one needs only the statistical property of the exponents instead of their detailed growth behavior. Because RMT behavior in the Lyapunov spectrum coincides with RMT behavior in the energy spectrum for the models we considered, the Lyapunov spectrum may be useful for connecting the different criteria for chaos. As a bonus, universality in the quantum Lyapunov spectrum has a classical counterpart \cite{Hanada:2017xrv}, so it may also be useful to connect classical and quantum chaos.

We emphasize that these universalities are merely empirical. There may be other observables that provide a similar characterization of quantum chaos which are also more accessible to experiment. In this paper, we consider time-ordered two-point correlators that are easier to study, both theoretically and experimentally, than OTOCs. Specifically, given a set of simple operators $\{O_j\}$ and their Heisenberg representations $O_j(t)=e^{i H t} O_j e^{-i H t}$, we consider the matrix of all possible two-point functions $\langle O_i(t) O_j(0)\rangle$, construct its time-dependent spectrum, and then study the statistical properties of the spectrum. Based on this study, we propose that this two-point correlation spectrum, which is roughly a spectrum of decay rates, has universal statistical properties for all chaotic systems.

Below, we first define the two models, SYK and XXZ, that we will consider. Next, we define a spectrum of decay rates derived from two-point functions and propose a universal behavior for the spectrum in chaotic systems. Then we provide detailed numerical evidence for the conjecture using finite size exact diagonalization studies.
\section{Models}

\subsection{SYK model}
The first example is the SYK model \cite{Maldacena:2016hyu,Sachdev:2015efa,Kitaev_talk} (see Ref.~\cite{Rosenhaus:2018dtp} for a recent review) consisting of $N$ Majorana fermions with  Hamiltonian
\begin{eqnarray}
\hat{H}
&=&
\sqrt{\frac{6}{N^3}}\sum_{i<j<k<l}J_{ijkl}\hat{\psi}_i\hat{\psi}_j\hat{\psi}_k\hat{\psi}_l
\nonumber\\
& &
\qquad
+
\frac{\sqrt{-1}}{\sqrt{N}}\sum_{i<j}K_{ij}\hat{\psi}_i\hat{\psi}_j.
\nonumber \\
\label{eqn:q=2deformedSYK}
\end{eqnarray}
Majorana  fermions satisfy the anti-commutation relations $\{\hat{\psi}_i,\hat{\psi}_j\}=\delta_{ij}$
and $J_{ijkl}$ is random Gaussian coupling with mean zero and standard deviation $1$. The Hamiltonian also includes a quadratic term, and $K_{ij}$ is Gaussian random with mean zero and standard deviation $K$. The dimension of the Hilbert space is $2^{N/2}$.
When $K=0$, this model is maximally chaotic at low temperatures, namely the Maldacena--Shenkar--Stanford bound \cite{Maldacena:2016hyu,Kitaev_talk} is asymptotically saturated.
When $K>0$, low-energy modes become non-chaotic, while high-energy modes remain chaotic~\cite{Garcia-Garcia:2017bkg,Nosaka:2018iat}.

\subsection{XXZ model with random magnetic field}
The second example is the XXZ model, a one-dimensional $S=1/2$ spin chain with random magnetic field along $z$-direction (see e.g.~\cite{PhysRevB.91.081103}),
\begin{eqnarray}
\hat{H}
=
\sum_{i=1}^{N_\mathrm{site}}\left(
\frac{1}{4}
\vec{\sigma}_i\vec{\sigma}_{i+1}
+
\frac{w_{i}}{2}\sigma_{z,i}
\right).
\label{eqn:XXZ}
\end{eqnarray}
Here $\vec{\sigma}=(\sigma_x, \sigma_y, \sigma_z)$ are Pauli matrices with periodic boundary condition $\vec{\sigma}_{N_{\rm site}+1}=\vec{\sigma}_1$.
The random magnetic fields $w_i$ are independent and uniformly distributed in $[-W,+W]$. At $W\gtrsim 3.5$, most of the energy eigenstates are in the many-body localized (MBL) phase \cite{PhysRevB.91.081103,serbyn1507criterion}. (For the physics of the MBL phase, see e.g.~\cite{PhysRev.109.1492,PhysRevLett.95.206603,BASKO20061126,aleiner:hal-00543657}.)

\section{Proposal and numerical evidences}
The starting point is choosing a set of operators and organizing the set of two-point functions into a matrix. The matrix of two-point functions, $G_{ij}^{(\phi)}(t)$, is defined by
\begin{eqnarray}
G_{ij}^{(\phi)}(t)
=
\langle
\phi\vert
\hat{\psi}_i(t)
\hat{\psi}_j(0)
\vert\phi
\rangle
\label{G-SYK}
\end{eqnarray}
for SYK, and by
\begin{eqnarray}
G_{ij}^{(\phi)}(t)
=
\langle
\phi\vert
\sigma_{+,i}(t)
\sigma_{-,j}(0)
\vert\phi
\rangle
\label{G-XXZ}
\end{eqnarray}
for XXZ, where $\sigma_{\pm}=\frac{\sigma_x\pm i\sigma_y}{2}$.
Here, we will take the state $\vert\phi\rangle$ to be an energy eigenstate.
Similar analysis can be done for other choices of $\vert\phi\rangle$, for example the spin diagonal states as we comment later.
Note also that we can consider other two-point functions, e.g.
$G_{ij}^{(\phi)}(t)=\langle\phi\vert\sigma_{z,i}(t)\sigma_{z,j}(0)\vert\phi\rangle$; the generalization to other systems is straightforward.

Let the singular values of $G_{ij}^{(\phi)}(t)$ be $e^{\lambda_i^{(\phi)}(t)}$.
We denote the $\lambda_i^{(\phi)}(t)$ as `exponents'. Our conjecture is two-fold:
\begin{itemize}
\item
In quantum chaotic systems, $G_{ij}^{(\phi)}$ becomes `random' at sufficiently large $t$ such that the exponents are described by RMT.

\item
In non-ergodic theories (e.g. the MBL phase) the exponents are not described by RMT.
\end{itemize}
The idea behind this conjecture is simple. When the system is chaotic, information about local perturbations
should be washed away. Hence it is natural to expect that $G_{ij}^{(\phi)}(t)$ becomes a random matrix for each realization of the Hamiltonian, choice of the state $|\phi\rangle$, and time $t$.
For example, if there is no particular symmetry, then $G_{ij}^{(\phi)}(t)$ should become a random complex matrix, and its singular values should follow Gaussian Unitary Ensemble (GUE).
On the other hand, if the system is not chaotic, some structure should survive and a deviation from RMT should be observable.
Note that while in Ref.~\cite{Gharibyan:2018fax} we used matrices of correlators that contain out-of-time-ordered terms, $G_{ij}^{(\phi)}(t)$ are purely time-ordered correlators.

It is unclear at present how this characterization of chaos is related to other characterizations, such as RMT universality in the energy spectrum or the exponential Lyapunov growth of OTOCs. The time evolution of the $N$-dimensional matrix $G_{ij}^{(\phi)}(t)$ reflects the unitary time evolution by the many-body Hamiltonian $\hat{H}$ in a nontrivial way.
Below, we demonstrate that these different characterizations are at least compatible in the SYK and XXZ models.

We note that the singular values of a general matrix $G$, which are the square roots of the (non-negative, real) eigenvalues of the matrix $G^\dag G$, are all non-zero if the rank of $G$ is equal to its dimension, but are not necessarily equal to the absolute values of the eigenvalues of $G$, because $G$ might not be a normal (\textit{i.e.} $G^\dag G = G G^\dag$) matrix.
We also note that other measures of spectral fluctuation and singular vector delocalization \cite{Berry1985PRSLA,Mueller2004PRL,Kos2018PRX} are also found to be compatible with RMT universality.
In the Supplemental Materials we present the number variance of the unfolded spectrum of the exponents as well as discuss the behavior of the singular vectors of the two-point-function matrix.

\subsection{Numerical study}
In this section, we calculate the exponents $\lambda_i^{(\phi)}(t)$ numerically and study their statistical features.
The exponents are sorted such that $\lambda_1^{(\phi)}(t)\ge \lambda_2^{(\phi)}(t)\ge\cdots\ge\lambda_N^{(\phi)}(t)$. The primary objects of study are the nearest-neighbor level spacing $s_i^{(\phi)}(t)\equiv \lambda_i^{(\phi)}(t)-\lambda_{i+1}^{(\phi)}(t)$ and the nearest-neighbor gap ratio $r_i=\frac{{\rm min}(s_i,s_{i+1})}{{\rm max}(s_i,s_{i+1})}$.

\subsubsection{Details of unfolding}
Because the number of exponents we can obtain numerically from each matrix of two-point functions is small,
we need to use the fixed-$i$ unfolding method \cite{Gharibyan:2018fax}: for each $i$, we rescale $s_i$ so that the average over many samples becomes 1.
Namely, we take $\tilde{s}_i=\frac{s_i}{\langle s_i\rangle}$,
where $\langle\ \cdot\ \rangle$ stands for the average over the samples (but not over different values of $i$).
We will call this $\tilde{s}_i$ simply as $s_i$ when there is no risk of confusion.
The values of $r_i$'s shown below are calculated from these rescaled $s_i$'s.

In order to reduce finite-$N$ effects further, we can use the sample-by-sample rescaling method as well. We shift the exponents so that the average becomes zero,
and then rescale them so that the standard deviation becomes 1. Namely,
$\lambda_i^{(\phi)}(t)\mapsto\tilde{\lambda}_i^{(\phi)}(t)=\alpha \lambda_i^{(\phi)}(t) + \beta$ so that $\sum_i \tilde{\lambda}_i^{(\phi)}(t) = 0$ and $\sum_{i=1}^{N} \tilde{\lambda}_i^{(\phi)}(t)^2 = N$.
This can remove the $N$-dependent fluctuation of the entire width.
When this method was applied to the energy spectrum of the SYK model \cite{Gharibyan:2018jrp}, the statistical properties were improved substantially.
In this paper, we have used only the fixed-$i$ unfolding for both the SYK model and the XXZ model. The sample-by-sample rescaling slightly changes the results, often bringing the results closer to random matrix or Poisson values, however we did not observe a qualitative change.

\begin{figure*}[htbp]
\begin{center}
\includegraphics{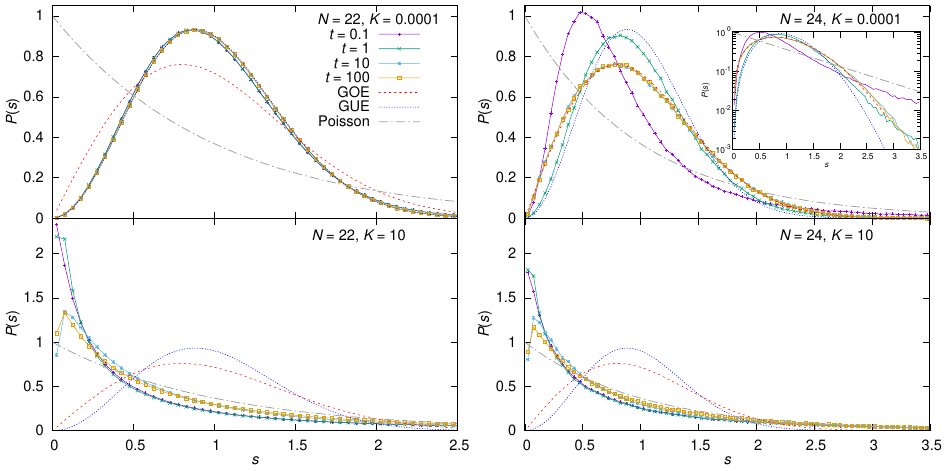}
\end{center}
\caption{
SYK, the distribution of nearest neighbor level spacing $P(s)$ for various values of $t$,
$K=0.0001$ and $K=10$.
All eigenstates are used and the larger $N/2$ exponents are used.  $N = 22, 24$.
Inset: log plot for $P(s)$ for $(N, K) = (24, 0.0001)$.
At short time the tail of the distribution is exponential, as in the case of the nearest neighbor spacing distribution for uncorrelated values (Poisson), while at longer times the tail agrees well with the distribution for GOE ($\sim s e^{-(\pi/4)s^2}$).
}\label{Fig:SYK-NN}
\end{figure*}

\begin{figure}[htbp]
\begin{center}
\includegraphics[width=7.85cm]{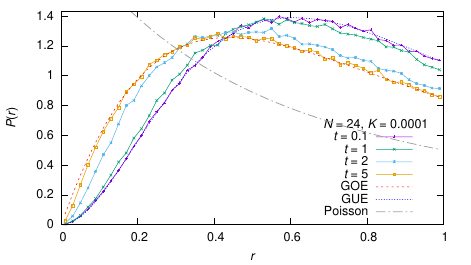}
\includegraphics[width=7.85cm]{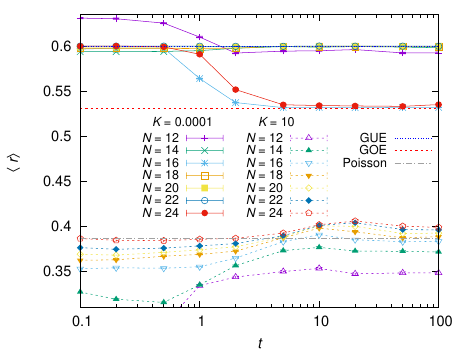}
\end{center}
\caption{
Upper: SYK, the distribution of nearest neighbor gap ratio $P(r)$ for $N = 24$, $K = 0.0001$.
Lower: SYK model, energy eigenstates, the averaged gap ratio as a function of the time $t$ for $N = 12, 14, \ldots, 24$ for $K=0.0001$ and $K=10$.
At least 2000 (16) samples are used for $N\leq 22$ ($N=24$).
}\label{Fig:SYK-r}
\end{figure}

\subsection{Symmetry of $G_{ij}^{(\phi)}$}
The matrix of two-point functions $G_{ij}^{(\phi)}$ is usually a complex matrix without particular symmetry.

For the XXZ model, $G_{ij}^{(\phi)}$ is complex and symmetric when the reference state is an energy eigenstate,
$|\phi\rangle=|E\rangle$.
To see this, first observe that the Hamiltonian is real and symmetric.
Hence $(e^{-iHt})^T=e^{-iHt}$. Also, the energy eigenstates can be chosen to be real unit vectors
(unless there is an accidental degeneracy in the energy spectrum, which does not happen for generic values of $w_i$'s),
and hence, $\langle E|\sigma_{+,i}(0)=(\sigma_{-,i}(0)|E\rangle)^T$.
Therefore,
\begin{eqnarray}
G_{ij}^{(\phi)}
&=&
\langle E|
\sigma_{+,i}(t)
\sigma_{-,j}(0)
|E\rangle
\nonumber\\
&=&
e^{iEt}
\langle E|
\sigma_{+,i}(0)
e^{-iHt}
\sigma_{-,j}(0)
|E\rangle
\nonumber
\end{eqnarray}
is complex and symmetric.

Next let us consider the SYK model.
When $|\phi\rangle$ is an energy eigenstate, then unless $K=0$ and $N\equiv 0\ (\mathrm{mod}~8)$ (namely $N$ is multiple of $8$), $G_{ij}^{(\phi)}(t)$ is a complex matrix without particular symmetry.
Hence, when the system is chaotic, if RMT behavior emerges, the relevant ensemble describing the singular values would be the Gaussian unitary ensemble (GUE).
When $K=0$ and $N\equiv0~(\mathrm{mod}~8)$, $G_{ij}^{(\phi)}(t)$ is complex and symmetric and in this case one expects Gaussian orthogonal ensemble (GOE) statistics. Hence we expect GOE when $K\simeq 0$ and $N\equiv 0\ (\mathrm{mod}~8)$.

The original Hamiltonian with $K=0$ exhibits different symmetry depending on the value of $N$ mod 8 --- GOE for $N\equiv 0$, GUE for $N\equiv 2, 6$ and GSE for $N\equiv 4$ \cite{Fu:2016yrv,You:2016ldz}.
For $N\equiv 0, 4\ (\mathrm{mod}~8)$, the Hamiltonian can be taken to be real and symmetric.
It can be seen by using the following representation:
\begin{eqnarray}
\hat{\psi}_1
&=&
\sigma_z\otimes\textbf{1}\otimes\textbf{1}\otimes\textbf{1}\otimes\cdots\otimes\textbf{1}\otimes\textbf{1},
\nonumber\\
\hat{\psi}_2
&=&
\sigma_y\otimes\sigma_y\otimes\sigma_y\otimes\sigma_y\otimes\cdots\otimes\sigma_y\otimes\sigma_y,
\nonumber\\
\hat{\psi}_3
&=&
\sigma_y\otimes\sigma_x\otimes\textbf{1}\otimes\sigma_y\otimes\cdots\otimes\sigma_y\otimes\sigma_y,
\nonumber\\
\hat{\psi}_4
&=&
\sigma_y\otimes\sigma_z\otimes\textbf{1}\otimes\sigma_y\otimes\cdots\otimes\sigma_y\otimes\sigma_y,
\nonumber\\
\hat{\psi}_5
&=&
\sigma_y\otimes\sigma_y\otimes\sigma_x\otimes\textbf{1}\otimes\cdots\otimes\sigma_y\otimes\sigma_y,
\nonumber\\
\hat{\psi}_6
&=&
\sigma_y\otimes\sigma_y\otimes\sigma_z\otimes\textbf{1}\otimes\cdots\otimes\sigma_y\otimes\sigma_y,
\nonumber\\
& &
\qquad\qquad\qquad
\cdots
\nonumber\\
\hat{\psi}_{N-3}
&=&
\sigma_y\otimes\sigma_y\otimes\sigma_y\otimes\sigma_y\otimes\cdots\otimes\sigma_x\otimes\textbf{1},
\nonumber\\
\hat{\psi}_{N-2}
&=&
\sigma_y\otimes\sigma_y\otimes\sigma_y\otimes\sigma_y\otimes\cdots\otimes\sigma_z\otimes\textbf{1},
\nonumber\\
\hat{\psi}_{N-1}
&=&
\sigma_y\otimes\textbf{1}\otimes\sigma_y\otimes\sigma_y\otimes\cdots\otimes\sigma_y\otimes\sigma_x,
\nonumber\\
\hat{\psi}_{N}
&=&
\sigma_y\otimes\textbf{1}\otimes\sigma_y\otimes\sigma_y\otimes\cdots\otimes\sigma_y\otimes\sigma_z.
\end{eqnarray}
When $N$ is a multiple of four, they are all real and symmetric.
(Note that $\sigma_x$ and $\sigma_z$ are real symmetric, while $\sigma_y$ is pure imaginary and anti-symmetric.)
Hence $\hat{H}=\sqrt{\frac{6}{N^3}}\sum_{i<j<k<l}J_{ijkl}\hat{\psi}_i\hat{\psi}_j\hat{\psi}_k\hat{\psi}_l$ is real,
and of course it is hermitian, and therefore, real and symmetric.
Note that $\hat{H}$ is not real when $K$ is not zero.

When $N\equiv 0\ (\mathrm{mod}~8)$ and $K=0$, the energy spectrum is not degenerate and the energy eigenstates are represented
by real vectors.
Therefore, just as in the case of the XXZ model, the matrix of two-point functions \eqref{G-SYK} is complex and symmetric for the energy eigenstates.
Hence we expect the GOE statistics.

The situation is a little bit complicated when $N\equiv 4\ (\mathrm{mod}~8)$, because the energy spectrum is two-fold degenerate.
In general, an energy eigenstate is not a real vector, but rather, just a linear combination of two real vectors with complex coefficients.
When we take $K$ to be small but nonzero, the degeneracy is split, and the energy eigenstates are generically far from real vectors.
Therefore, we expect the GUE statistics.

We can also see that $G_{ij}^{(\phi)}(t=0)$ is real and symmetric when $N\equiv 0\ (\mathrm{mod}~8)$ and $K=0$. From this it follows that
$G_{ij}^{(\phi)}(0)=\frac{G_{ij}^{(\phi)}(0)+G_{ji}^{(\phi)}(0)}{2}=\frac{1}{2}\langle E|\{\hat{\psi}_i,\hat{\psi}_j\}
|E\rangle=\frac{\delta_{ij}}{2}$.
Namely all exponents are $-\log 2$ at $t=0$.

\subsection{Numerical results}

At the values of $N$ we study, the energy dependence of the spectrum is not large. (The energy dependence is similar to the case of the Lyapunov spectrum; see Ref.~\onlinecite{Gharibyan:2018fax} for a detailed explanation.) Hence, it is simplest to average over all energy eigenstates. Numerically we find that the gap between $\lambda_{N/2}$ and $\lambda_{N/2+1}$ is bigger than the other gaps and appears to behave differently when $K$ is large, as observed in the plots in Sec.~\ref{sec:time-dep-lambda-i}. Hence, we use only the first half of the spectrum with $N/2$ exponents in the analysis. We checked that similar results are obtained using the other half of the spectrum.

Fig.~\ref{Fig:SYK-NN} shows the nearest-neighbor level spacing for the SYK model with $N=22$ and $24$. For $N=22$, near $K=0$ (chaotic phase) the spectrum is GUE-like~\footnote{A small nonzero value $K=0.0001$ is used to avoid a degeneracy in the energy spectrum, that leads to ambiguity in the choice of eigenstates, when $K$ is exactly zero.}. It is interesting that the GUE behavior can be seen at all time scales. We observed the same phenomenon for other $N\not\equiv 0\ (\mathrm{mod}~8)$.

For $N\equiv 0\ (\mathrm{mod}~8)$, as in the case of $N=24$, the spectrum is GOE-like at sufficiently late time, but at early time there are large deviations from GOE. In the opposite limit of large $K$, in which the system is not chaotic to leading order, the spectrum is Poisson-like. This claim is substantiated in Fig.~\ref{Fig:SYK-r} which shows the nearest-neighbor gap ratio $r$.
In the upper panel, we plot the distribution of $r$ for $t=0.1, 1, 2, 5$ for $N=24$ and $K=0.0001$. At early times $P(r)$ agrees with GUE, while   as we have seen in Fig.~\ref{Fig:SYK-NN}, $P(s)$ does not. We do not think that this is related to universality at the sufficiently late times, because there is no apparent reason that GUE should appear. Note also that, because the number of exponents is small, it is natural to expect that $P(r)$ suffers from more significant finite-size correction than $P(s)$. For $t\gtrsim 5$, $P(r)$ for $N=24$ agrees well with GOE.

In the lower panel of Fig.~\ref{Fig:SYK-r}, we plot the average of $r$, $\langle r\rangle$, against time.
At late times, for $K\simeq 0$, the values converge to those of GUE or GOE as expected from the symmetry at each $N$, namely GUE for $N\not\equiv 0\ (\mathrm{mod}~8)$ and GOE for $N\equiv 0\ (\mathrm{mod}~8)$.
For large $K$, $\langle r\rangle$ stays close to the Poisson value.
We discuss the $N$-dependence of the behavior of $\langle r\rangle$ later.

\subsection{XXZ model}
\begin{figure}[htbp]
\begin{center}
\includegraphics[width=7.85cm]{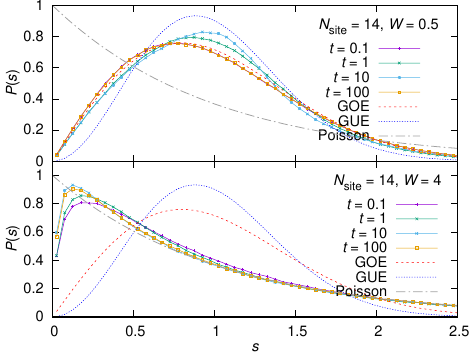}
\end{center}
\caption{
The distribution of nearest-neighbor level spacing $s$, XXZ, $t=0.1,10,20,100$ for $W=0.5$ and $W=4$,
with $N_{\rm site}=14$ for central $10~\%$ of the energy eigenstates.
The largest $N_\mathrm{site}/2$ exponents are used.
}\label{Fig:XXZ-NN-E}
\end{figure}

\begin{figure}
\begin{center}
\includegraphics[width=8cm]{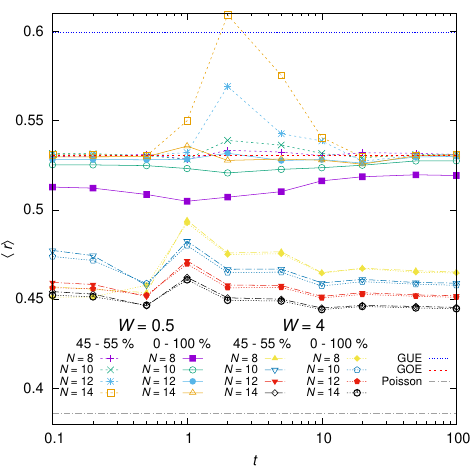}
\end{center}
\caption{
The averaged nearest-neighbor gap ratio for the central $10~\%$ ($45-55\%$) and all ($0-100\%$) of the energy eigenstates.
At least 22000 (1200) samples are used for $N_\mathrm{site}=8, 10, 12$ ($N_\mathrm{site}=14$).
The largest $N_\mathrm{site}/2$ exponents are used.
In the figure, $N_{\rm site}$ is shown as $N$ for brevity.
}\label{Fig:XXZ-r-E}
\end{figure}

Now consider the XXZ model. This model conserves the $z$-component of the total spin, and we consider only the $S^z=0$ sector. We study two values of the $W$ parameter, $W=0.5$ (the ergodic phase) and $W=4$ (the MBL phase). In this model, $G^{(\phi)}$ is complex and symmetric when $|\phi\rangle$ is an energy eigenstate. Hence, in the ergodic phase, we expect GOE statistics for the singular values.

We first discuss the time-scale for the decay of the two-point functions. For $W=4$, we observe a clear split of the upper and lower halves.
Hence, the larger half of the exponents is used for the analysis, both for $W = 0.5$ and for $W = 4$. We checked that the result does not change much if the lower half, or all the exponents, are used provided $N_{\rm site}$ is large enough ($N_{\rm site}=12,14$).

The energy dependence is rather large unlike the SYK model. (Again, see Ref.~\onlinecite{Gharibyan:2018fax} for a detailed explanation.) Hence we need to restrict the energy to be in a small range in order to remove an uncontrolled energy variation from the analysis.

Fig.~\ref{Fig:XXZ-NN-E} shows the distribution of the nearest-neighbor level spacing. The chaotic phase exhibits a GOE distribution, while the distribution is close to Poisson in the MBL phase. Note that, unlike the SYK model, the chaotic phase is not described by GOE at early time. Interestingly, the deviation from GOE becomes large at $1\lesssim t\lesssim 10$, but it eventually vanishes~\footnote{The agreement is improved to some extent by removing the largest exponent, which decays much slower than others. Still, the deviation at short time remains.}.
There is a curious $N_{\rm site}$-dependence of this deviation at intermediate time which is discussed further in Sec.~\ref{subsec:Nsite-XXZ}, also with the data for $W = 1, 2, 3$.

In Fig.~\ref{Fig:XXZ-r-E} the averaged nearest-neighbor gap ratio is plotted, both for the central 10~\% and for all of the energy spectrum. In the chaotic phase ($W=0.5$), the value of $\langle r\rangle$ does not depend on $t$ at late time and approaches the GOE value \cite{Atas}. The agreement with the GOE value even at intermediate time for small $N_\mathrm{site}$ or for the entire spectrum is likely a coincidence. For the center of the spectrum, as $N_\mathrm{site}$ is increased, the value of $\langle r\rangle$ as well as the nearest-neighbor level separation $P(s)$, shown in the upper panel of Fig.~\ref{Fig:XXZ-NN-E}, deviate from those for GOE. In the MBL phase ($W=4$), $\langle r\rangle$ is smaller than the GOE value and decreases toward the Poisson value as $N_\mathrm{site}$ increases, both for the central 10~\% and for all of the energy spectrum, as we discuss in more detail later.

\subsection{Time dependence of the exponents $\lambda_i$}\label{sec:time-dep-lambda-i}

\begin{figure}
\includegraphics[width=8cm]{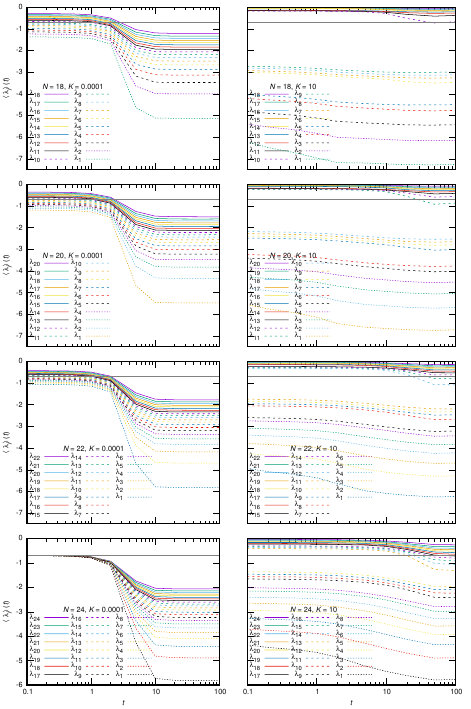}
\caption{SYK model, energy eigenstates, averaged exponents for $K=0.0001$ (left), $K=10$ for $N=18,20,22,24$ from top to bottom as functions of time $t$. The horizontal gray line indicates $-\log 2$.
}
\label{fig:SYK2Pv2-N18-24-bare}
\end{figure}

\begin{figure}
    \centering
    \includegraphics[width=8cm]{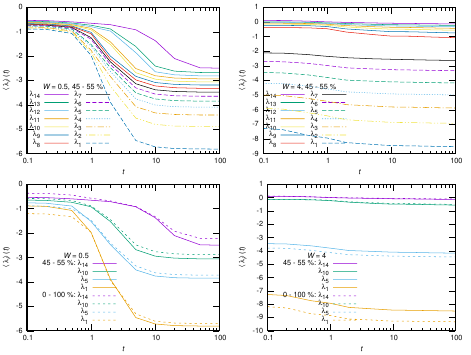}
    \caption{XXZ model, energy eigenstates, averaged exponents for $N_\mathrm{site}=14$ and $W=0.5$ (left), $4$ (right) as functions of time $t$.
Averages for the central 10\% of the eigenstates (45 - 55 \%), shown in the upper row, and all the eigenstates (0 - 100\%) are compared in the plots in the lower row for $\lambda_1, \lambda_5, \lambda_{10}, \lambda_{14}$.
}
    \label{fig:XXZ2pEig-N14-lambdaj}
\end{figure}

For the SYK model, in Fig.~\ref{fig:SYK2Pv2-N18-24-bare}, the exponents $\lambda_i(t)$ are plotted as functions of time $t$ for $K=0.0001, 10$ and various $N$.
(When $N\equiv 2$ or 6, it is possible to uniquely specify the energy eigenstates by taking into account parity as well \cite{Fu:2016yrv,You:2016ldz}.)
Note that, when $N\equiv 0$ and $K\simeq 0$, all exponents are close to $-\log 2$ at early time, as we have already explained.
Presumably, this is the reason for the different early-time behaviors in the level statistics of $N\equiv 0$ and $N\equiv 2,4,6$.

Next let us consider the XXZ model. In Fig.~\ref{fig:XXZ2pEig-N14-lambdaj}, the exponents $\lambda_i(t)$ are plotted as functions of time $t$, for $W=0.5, 4$ and $N_{\rm site}=14$.
In the top panels, we have plotted all the exponents, by using 10\% of the energy eigenstates in the middle of the spectrum.
For $W=0.5$, the largest exponent decays more slowly (decay time scale $\sim 20$) than the other exponents (decay time scale $\sim 10$).
Due to this, a better agreement with GOE can be seen when we remove the largest exponent from the analysis of the statistical property.
(Even with the largest exponent, the agreement with GOE is still good.)
For $W=4$, we can see a clear split of the upper and lower halves.
Hence we use the larger half for the analysis of the statistical property.
(The result does not change much when we use the lower half, or all exponents, at sufficiently large $N_{\rm site}$ ($N_{\rm site}=12,14$).)
In the bottom panels, we have compared the exponents calculated by using 10\% of the energy eigenstates in the middle of the spectrum,
and all the energy eigenstates. There are visible differences, and due to them, better agreement with GOE can be seen when the energy range is restricted.

\subsection{Dependence on $N$ for the SYK model}

\begin{figure}
\includegraphics[width=8cm]{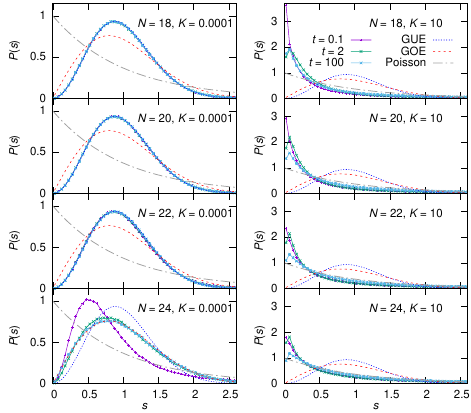}
\caption{
The distribution of nearest-neighbor level spacing $s$, SYK, $t=0.1,2,10$ for $K=0.0001$ and $K=10$,
with $N=18, 20, 22$ and $24$.
The larger $N/2$ exponents are used.
}\label{fig:SYK-N18-24-NN-E}
\end{figure}

In Fig.~\ref{fig:SYK-N18-24-NN-E}, the nearest neighbor level spacing $s$ for the SYK model is plotted,
for various values of $N$. When $K$ is close to zero, GUE and GOE can be seen at sufficiently late time,
for $N\equiv 2,4,6$ and $N\equiv 0$ mod 8, respectively.
Early-time behaviors are rather different: for $N\equiv 2,4,6$, GUE can be seen from $t=0$,
while for $N\equiv 0$ substantial deviation from GOE can be seen. Presumably this deviation is related to the exact degeneracy
of the exponents at $t=0$.
When $K$ is large, we do not see RMT at all.
The nearest gap ratio $\langle r\rangle$ plotted in the lower panel of Fig.~\ref{Fig:SYK-r} shows the same pattern.

\subsection{Dependence on $N_{\rm site}$ for the XXZ model}
\label{subsec:Nsite-XXZ}

\begin{figure}
\includegraphics[width=8cm]{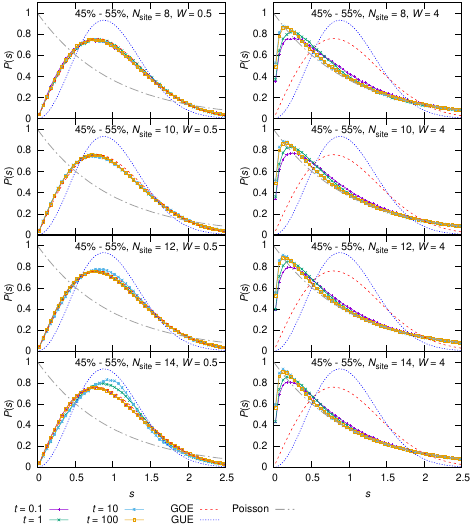}
\caption{
The distribution of nearest-neighbor level spacing $s$, XXZ, $t=0.1,10,20,100$ for $W=0.5$ and $W=4$,
with $N_{\rm site}=8, 10, 12$ and $14$.
The larger $N_\mathrm{site}/2$ exponents, for eigenstates coming within the 45\% - 55\% range when sorted by the energy, are used.
Rescale and shift method is not used, only the fixed-$i$ unfolding has been used.
}\label{Fig:XXZ-N8-14-NN-E}
\end{figure}

\begin{figure}
\includegraphics[width=8cm]{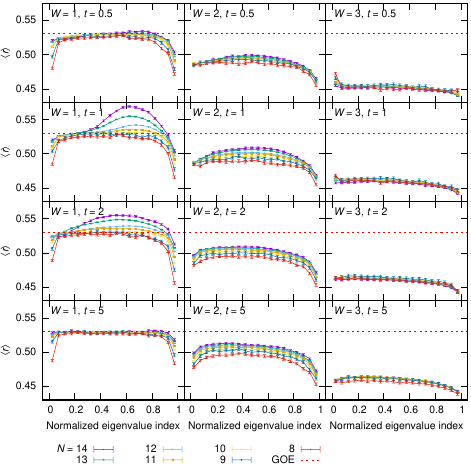}
\caption{Dependence of $\langle r\rangle$ on normalized eigenstate energy index for $N_\mathrm{site} = 14, 13, \ldots, 8$ and $W = 1, 2, 3$ at difference times.
The average over eigenstates in the $5~\%$ bins are plotted.}
\label{fig:XXZ2pEig-N14-8-index-dependence}
\end{figure}

\begin{figure}
\includegraphics[width=8cm]{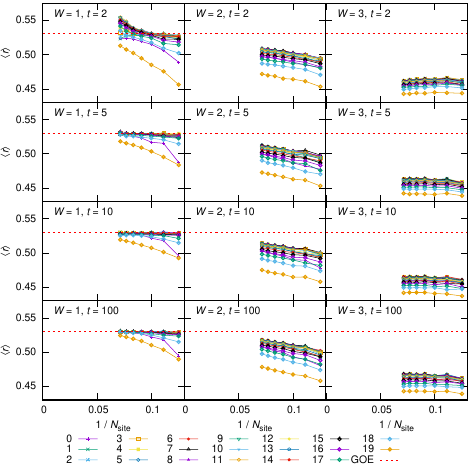}
\caption{Dependence of $\langle r\rangle$ on $1/N_\mathrm{site}$ for $N_\mathrm{site} = 14, 13, \ldots, 8$ and $W = 1, 2, 3$ at difference times,
averaged for the 20 bins according to the eigenstate energy indicated by $0\ldots 19$.
The average over eigenstates in the $5~\%$ bins are plotted.}
\label{fig:XXZ2pEig-N14-8-invN-dependence}
\end{figure}

In Fig.~\ref{Fig:XXZ-N8-14-NN-E}, we have plotted the distribution of the nearest neighbor level spacing for $N_{\rm site}=8, 10, 12$ and $14$.
We can see the GOE distribution in the chaotic phase at late time ($t=100$ in the plots), while the distribution is close to Poisson in the MBL phase.
Note that the chaotic phase is not described by GOE at early time.
Indeed, at $t=0.1$, we can see small but non-negligible deviation from GOE.
Interestingly, the deviation at intermediate times becomes larger at larger $N_{\rm site}$.

In Fig.~\ref{fig:XXZ2pEig-N14-8-index-dependence}, we have plotted the dependence of the nearest neighbor gap ratio on the eigenstate energy for $W = 1, 2, 3$ at various times.
In Fig.~\ref{fig:XXZ2pEig-N14-8-invN-dependence} the same data is plotted against $1/N$.
We observe that the quantity reaches its late time value around $t = 10$.
For $W = 1$ and $2$, the $N\to\infty$ value of $\langle r\rangle$ converges close to the GOE value for the majority of the energy index range,
though the deviation close to the edge of the spectrum is larger for $W = 2$, which is consistent with the phase diagrams in the literature on this model \cite{PhysRevB.91.081103,serbyn1507criterion} (close to the edge of the energy spectrum, the MBL phase appears).
For $W = 3$, the quantity exhibits weaker time dependence, and it no longer increases as $N$ is increased.

\section{Summary and Discussion}
Here we introduced a spectrum defined from a matrix of two-point functions
(\eqref{G-SYK} for SYK and \eqref{G-XXZ} for XXZ), and proposed that the statistical features of this spectrum
exhibit random matrix universality when the underlying system is chaotic.

While we have used the energy eigenstates to define the spectrum, this particular choice is not crucial to observe universality. Spin eigenstates such as $\vert\!\uparrow\uparrow\cdots\uparrow\uparrow\rangle$ and $\vert\!\uparrow\downarrow\cdots\uparrow\downarrow\rangle$ also yield the same structure at long time~\cite{in-preparation}, but the time-scale for the onset of RMT-behavior can depend on the choice of state.

In this paper, all the models considered have some degree of disorder in their definition. One could worry that this disorder is the source of the RMT behavior. The fact that we do not observe RMT signatures in the MBL phase shows that this is not so.
Also we note that we have used the fixed-$i$ unfolding method and limited our analysis to the short-range spectral fluctuation because the number of exponents accessible with exact diagonalization is not large.

There are various generalizations and extensions of this work. One clear task is to see if the same signatures are observed in other chaotic models. Another goal is an analytic argument for the observed behavior. Besides exact diagonalization, numerical methods such as the time-dependent density-matrix renormalization group may be employed to access two-point functions. This may allow one to study larger systems. Also, studying sub-matrices by limiting the spatial range of operators may give insights into the time scale for the onset of chaotic behavior, which may be important in studying operator spreading and information scrambling. If RMT universality can be observed there, it would provide a powerful tool to study the chaotic nature of large systems where real-time dynamics can be hard to access numerically. Also, while neither of the models we have studied here is with known classical behavior, the matrix of two-point functions considered here can be defined in classical systems as well. Whether the same universality can be found in that context is another interesting question.

\section*{Acknowledgments}
We thank S.~Hikami, S.~Matsuura and H.~Shimada for stimulating discussions.
This work was partially supported by JSPS KAKENHI Grants 17K14285 (M.~H.), 17K17822 (M.~T.), and JP20K03787 (M.~T.), the Simons Foundation via the It From Qubit Collaboration (B.~S.), and the Department of Energy award number DE-SC0017905 (B.~S).
H.~G. was supported in part by NSF grant PHY-1720397.
M.~H. thanks Brown University for the hospitality during his stay while completing the paper, and acknowledges the STFC Ernest Rutherford Grant ST/R003599/1.

\bibliography{universality}
\newpage

\setcounter{page}{1}

\begin{widetext}
\begin{center}
{\huge
\textbf{Supplemental material for ``A characterization of quantum chaos by two-point correlation function''}
}
\end{center}
\end{widetext}

\setcounter{section}{0}
\section{The number variance of the unfolded spectrum} \label{appendix:Sigma2}

\begin{figure*}
\includegraphics[width=11cm]{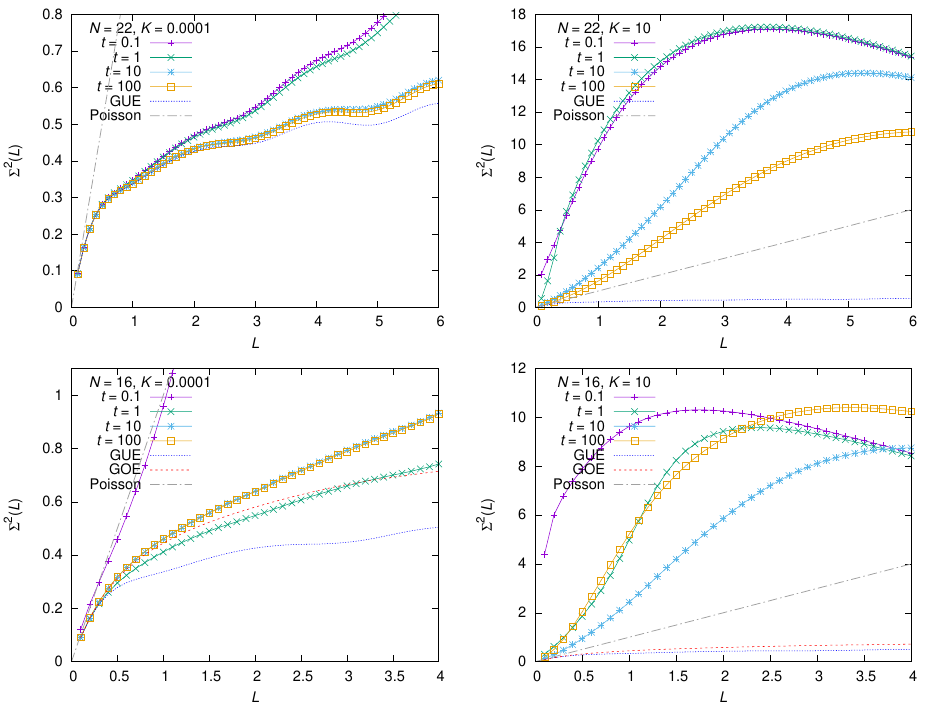}
\caption{SYK model, the number variance $\Sigma^2(L)$ plotted against the range $L$ of the unfolded spectrum for $G_{ij}^{(\phi)}=\langle\phi\vert\psi_{i}(t)\psi_{j}(0)\vert\phi\rangle$ with $N=22$ (top) and $N=16$ (bottom). 1000 samples have been used.
The data for $N$-dimensional GOE and GUE matrices are obtained using $10^7$ samples.}
\label{fig:SYK2P-Sigma2-N22-16}
\end{figure*}

\begin{figure*}
\includegraphics[width=11cm]{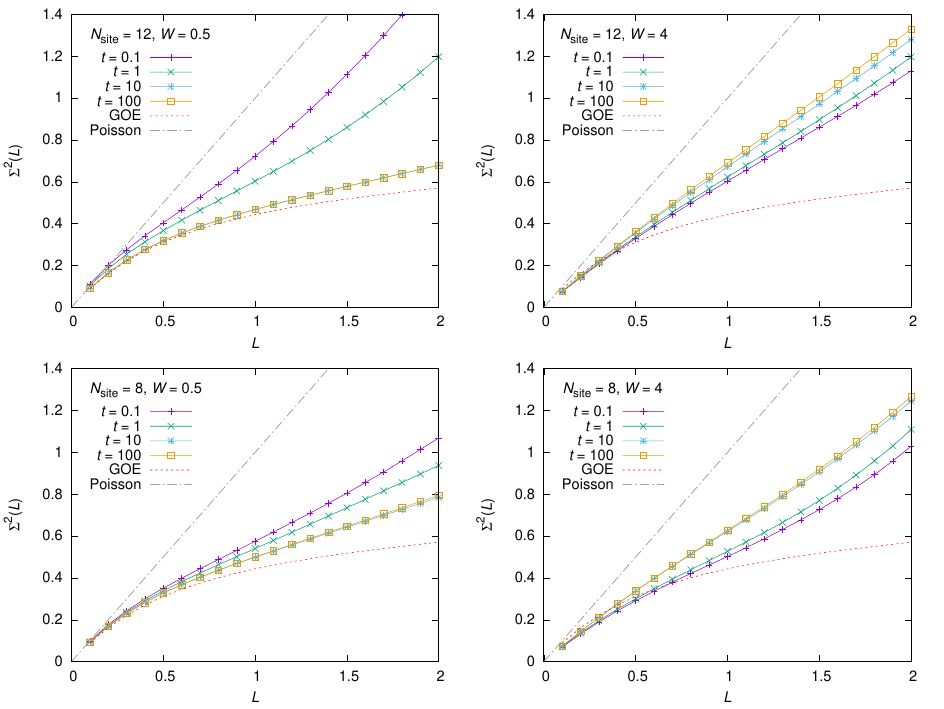}
\caption{XXZ model, the number variance $\Sigma^2(L)$ plotted against the range $L$ of the unfolded spectrum for $G_{ij}^{(\phi)}=\langle\phi\vert\sigma_{+,i}(t)\sigma_{-,j}(0)\vert\phi\rangle$ with $N_\mathrm{site}=12$ (top) and $N_\mathrm{site}=8$ (bottom). At least 20000 samples have been used.
The data for $N_\mathrm{site}$-dimensional GOE matrices are obtained using $10^7$ samples.}
\label{fig:XXZ2PEig-SpSm-Sigma2-N12-8}
\end{figure*}

In this section we present another indicator of the random matrix behavior of the exponents obtained from the singular values of the correlation functions.
The number variance $\Sigma^2(L)$ is the variance of the number of unfolded exponents in the section of spectrum of width $L$, where $L$ unfolded exponents are expected on average.
It is equal to $L$ for the uncorrelated case, while it grows logalithmically as a function of $L$ for random matrix ensembles \cite{mehta2004random}.
Here, we compare the values of $\Sigma^2(L)$ against those calculated for the eigenvalues of the $N$ (for the SYK model) or $N_\mathrm{site}$ (for the XXZ spin chain) dimensional random matrices.
The spectra are unfolded so that the average spacing is unity and are shifted so that the average becomes zero, then the variance of number of the values in the unfolded and shifted spectra within $[-L/2, L/2]$ are obtained.

For comparison to the SYK model, in Fig.~\ref{fig:SYK2P-Sigma2-N22-16}, the largest $N/2$ eigenvalues of $N/2$-dimensional symmetric real (GOE) and hermitian complex (GUE) random matrices are used for calculating the number variance.
We observe that for $N=22$ and $K=0.0001$, the results at $t=10$ and $100$ strongly resemble that for the GUE random matrix, including the finite-scale fluctuations.
The result for $N=16$ and $K=0.0001$ resembles that for the GOE random matrix at long times as expected.
On the other hand, for $K=10$, both for $N = 22$ and for $N = 16$ we observe larger variances than expected for the uncorrelated case (``Poisson'') in the plots, presumably because the exponents coalesce with each other.

In Fig.~\ref{fig:XXZ2PEig-SpSm-Sigma2-N12-8}, we compare the number variance between the exponents obtained from the singular values of eq.~\ref{G-XXZ} for the XXZ model and the eigenvalues of GOE random matrices with $N_\mathrm{site}=12$ and $N_\mathrm{site}=8$.
In both cases, half of the spectrum is unfolded and shifted for the computation of $\Sigma^2(L)$.
For $W=0.5$ in the chaotic regime, the increase of $\Sigma^2(L)$ as a function of $L$ is slowed down as $t$ is increased and almost converged for $t\gtrsim 10$ near the GOE result.
For $W=4$ in the many-body localized regime, the increase is closer to linear and to the result for the Poisson distribution, $\Sigma^2(L)=L$.

\section{Inverse participation ratio for the singular vector} \label{appendix:IPR}
\begin{figure}
\includegraphics[width=8cm]{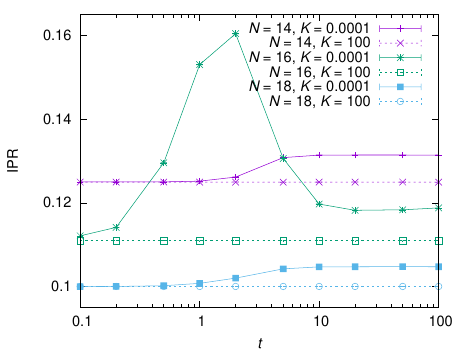}
\includegraphics[width=8cm]{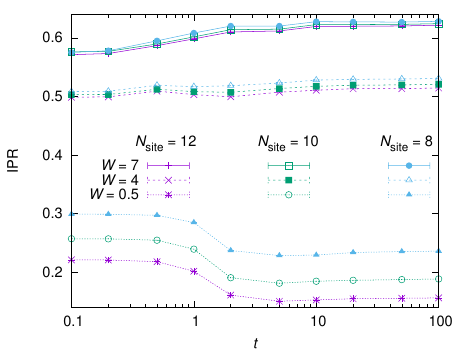}
\caption{The inverse participation ratio calculated for the right singular vectors of the correlation functions.
Top: eq.~\eqref{G-SYK} for the SYK model.
Bottom: eq.~\eqref{G-XXZ} for the XXZ model.
}
\label{fig:IPR-SYK-XXZ}
\end{figure}

The inverse participation ratio (IPR) for a wavefunction expressed on a basis $\{\vert j\rangle\}$, $\vert\psi\rangle = \sum_j \psi_j\vert j\rangle$, is defined as
\begin{equation}
\mathrm{IPR}(\vert \psi\rangle) = \frac{\sum_j\vert \psi_j\vert^4}{\left(\sum_j\vert \psi_j\vert^2\right)^2},
\end{equation}
in which the denominator is unity when the wavefunction is normalized ($\langle j\vert j \rangle = \sum_j \vert \psi_j\vert^2 = 1$).
In this definition, when the dimension of the Hilbert space is $D$, the IPR takes its smallest value, $1/D$, when all $D$ components of the wavefunction has the equal absolute value and thus most delocalized in the basis, $\vert\psi_j\vert = 1/\sqrt{D}$. A more localized wavefunction has a larger IPR, and in the most localized case with only one component having nonzero absolute value, $\vert \psi_{j_0}\vert = 1$, the IPR equals unity.
Thus the IPR parametrizes the localization of the wavefunction in the given basis and can be regarded as the inverse of the number of components having significant amplitude.

The IPR can also be calculated for the (left and right) singular vectors of the correlation matrix that we compute in our work.
We have checked that the values from the left and right singular vectors are indistinguishable. Therefore, we have used the IPR for the right singular vectors in the following. In Fig.~\ref{fig:IPR-SYK-XXZ} we plot the average of the IPR over all the right singular vectors obtained for the correlation function matrix for all energy eigenstates as a function of the time.

For the SYK model (eq.~\eqref{G-SYK}), the average of the IPR is small through the time evolution and decreases as $\sim 1/N$ as $N$ is increased as shown in the top panel of Fig.~\ref{fig:IPR-SYK-XXZ}.
For the almost integrable Hamiltonian with the large $K=100$, the value is almost independent of the time.
For the maximally chaotic case with very small $K$, for $N=14$ and $18$ the value slightly increases during $1\lesssim t \lesssim 10$ and stabilizes, while it has a peak for $N=16\equiv 0$ mod $8$, presumably correlated with the change of the symmetry class (see Fig.~\ref{Fig:SYK-r}). The small IPR does not guarantee the random matrix behavior, and the behavior observed here is consistent with the results of other quantities presented in the main text.

For the XXZ model (eq.~\eqref{G-XXZ}), for large $W$ corresponding to the many-body localized region, the IPR is large and not significantly changed as a function of $t$ or $N_\mathrm{site}$. The singular vectors of the matrix $G_{ij}^{(\phi)}(t)$ are also strongly localized, as in the bottom panel of Fig.~\ref{fig:IPR-SYK-XXZ}.
On the other hand, for $W = 0.5$, the IPR decreases as a function of $t$ before stabilizing at a value that is almost in proportion to $1/N_\mathrm{site}$, indicating the delocalized nature of the singular vectors.

\end{document}